# A study of stability of the "plasma - beam" system for the Lorentz plasma

## *Iogann Tolbatov*

*Mulliken Center for Theoretical Chemistry, Institute for Physical and Theoretical Chemistry, University of Bonn, Bonn, Germany (Emails: tolbatov.i@gmail.com)*

**Abstract**

The stability of the system comprising the cold immobile Lorentz plasma of density $N_e$ and the directed nonrelativistic (velocity $\vec{V}$) electronic beam of a small density $N'_e \ll N_e$ is investigated. The instability increment of the system is found via analysis of the electric permittivity $\varepsilon(\omega)$ in the limiting case of $\omega$ being much greater than the effective frequency of collisions. A conclusion is made that the system in question is stable with respect to oscillations with relatively small values of $\vec{k}\vec{V}$.

## 1. Introduction

As the kinetic coefficients of plasma are calculated, both the electron-electron- (*ee*-) and electron-ion- (*ei*-) collisions are generally to be taken into account. However, the role of the *ei*-collisions becomes predominant when the ion charges are high. Indeed, the frequency of these collisions $\varrho_{ee}$ as well as the cross-section are proportional to $(e^2)^2$. The frequencies of the *ee*- and *ei*- collisions are proportional to $N_e$ and $(ze^2)^2 N_i = z_i e^4 N_e$, respectively. Thus, as $z \gg 1$, the frequencies become $\varrho_{ei} \gg \varrho_{ee}$. Lorentz plasma is the plasma in which the *ee*-collisions can be considered neglectable relative to the *ei*-collisions [1].

The Lorentz plasma case is very interesting in methodological sense. Moreover, it can be applied to the weakly ionized gas, where the collisions of electrons with the neutral atoms take place instead of the *ei*-collisions [1]. The Lorentz plasma is an important notion used in the general formulation of plasma density and effective collision frequency of lowly and highly collisional plasmas [2] and electromagnetically induced transparency of atomic systems [3].

Lorentz plasma is an object of multiple investigations. A steady state heat flow in a Lorentz plasma as well as both the appropriate collisionless and collision dominated limits were studied by a discrete ordinate method [4]. The damping in the Lorentz plasma was investigated analytically by Comisar. It was found that superposition of collisional and Landau damping is taking place [5]. Later a numerical simulation confirmed this [6]. The phenomenon of the runaway electrons in an ideal Lorentz plasma was investigated by a variety of methods [7, 8] as well as the electrons suppression in a weak uniform electric field in a fully ionized Lorentz plasma [9, 10].

A very important aspect is the fact that the Lorentz plasma model is highly appropriate for studying transport coefficients such as electric conductivity, since the electron-electron collisions do not contribute to conductivity. The electron distribution function for a Lorentz plasma in a strong magnetic field was calculated by Radin [11]. The magnetic field was assumed as strong enough for a significant effect on the collision process, however not dominating the plasma completely. This presupposition allowed the computation of the accurate thermoelectric transport coefficients and viscosities [11]. The nonlocal *dc*-electrical conductivity of a Lorentz plasma in a stochastic magnetic field was studied by Jacobson *et al.* [12]. The transport equation and macroscopic laws of Lorentz plasma without magnetic field were used to investigate the transport coefficients in Lorentz plasma with the power-law $\kappa$-distribution. Expressions for thermoelectric coefficient, thermal conductivity, and electric conductivity were derived accurately [13]. The explicit quantum-mechanical expressions for the Lorentz plasma conductivity and resistivity tensors in a magnetic field were obtained recently. It was shown also that this model is applicable to plasmas with finite-mass ions when electron and ion temperatures are approximately equal [14].

The Lorentz plasma model has its limitations. It is only reasonably applicable for plasma with highly ionized ions with $Z \geq 10$. Although in the case of plasmas with $Z < 10$ the electron-electron collisions do not contribute directly to the induced current density due to the momentum conservation $(\int \vec{v} J_{ee}[f] d\vec{v} = 0$, they influence the permittivity value through the modified electron distribution function [15].

Our goal is to study the instability of a directed electron beam in the immobile plasma. We assume that the sum of electron charge densities in plasma and beam is equal to the ionic charges density in plasma. We consider a uniform and unbounded system filling up the whole space. The directed nonrelativistic velocity $\vec{V}$ of the beam is the same everywhere.

## 2. Calculational Details

After formulation of the effective frequency of collisions $\varrho_{ei}(v) = N_i v \sigma_t^{(ei)}$, where $\sigma_t^{(ei)} = \frac{4\pi z^2 e^4 L}{m^2 v^4}$ is the transport cross-section of the electron scattering on ions, one can introduce it in the $-i\omega \vec{P} = \vec{j}$ formulation of the dielectric permittivity. After averaging of $\vec{v}$ along directions, we obtain:

$$\varepsilon(\omega) = 1 - \frac{4\pi e^2}{3\omega T} \int \frac{v^2 f_0 d^3 p}{\omega + i\varrho(v)}. \tag{1}$$

In the limiting case $\omega \gg \varrho(v_T) = \frac{4\pi z e^4 N_e L}{m^{\frac{1}{2}} T_e^{\frac{3}{2}}}$, the dielectric permittivity becomes:

$$\varepsilon(\omega) = 1 - \frac{4\pi e^2 N_e}{m \omega^2} + i \frac{4\pi e^2 N_e}{3\omega^3 T} \langle v^2 \varrho(v) \rangle. \tag{2}$$

After the Maxwell averaging, we obtain:

$$\varepsilon(\omega) = 1 - \frac{\Omega_e^2}{\omega^2} + i \xi \frac{\Omega_e^3}{\omega^3}, \tag{3}$$

where $\xi = \frac{2\sqrt{2}}{3} \frac{z e^3 L \sqrt{N_e}}{T^{\frac{3}{2}}}$ and $\Omega_e = \sqrt{\frac{4\pi e^2 N_e}{m}}$ [1]. (4)

We assume that the beam and the plasma are cold, i.e. the thermal movement of particles is neglected. In the region of the electronic oscillations, the dielectric permittivity of the "plasma - beam" system is:

$$\varepsilon(\omega, \vec{k}) - 1 = -\frac{\Omega_e^2}{\omega^2} + i\xi \frac{\Omega_e^3}{\omega^3} - \frac{\Omega'_e{}^2}{\omega'^2} + i\xi' \frac{\Omega'_e{}^3}{\omega'^3}. \tag{5}$$

The first two terms relate to the immobile plasma, the third and the fourth – to the beam electrons. The beam's frame of reference $K'$ moves with the beam. $N'_e$ is the density of electrons in the beam. The parameters $\xi'$ and $\Omega'_e$ differ from their plasma system of reference counterparts (4) by inclusion of the $N'_e$ instead of $N_e$, whereas the frequency becomes $\omega' = \omega - \vec{k}\vec{V}$. Thus, the expression (5) transforms into:

$$\varepsilon(\omega, \vec{k}) - 1 = -\frac{\Omega_e^2}{\omega^2} + i\xi \frac{\Omega_e^3}{\omega^3} - \frac{\Omega'_e{}^2}{(\omega - \vec{k}\vec{V})^2} + i\xi' \frac{\Omega'_e{}^3}{(\omega - \vec{k}\vec{V})^3}. \tag{6}$$

We consider the density of the electronic beam to be small $N'_e \ll N_e$, yielding thus $\Omega'_e \ll \Omega_e$. Then the appearance of the beam only slightly changes the main branch of the plasma oscillation spectrum. Thus if $\varepsilon(\omega, \vec{k}) = 0$, then $\omega \approx \Omega_e$. From the equation (6), we obtain then:

$$\frac{\Omega_e^2}{\omega^2} + \frac{\Omega'_e{}^2}{(\omega - \vec{k}\vec{V})^2} - i\xi \frac{\Omega_e^3}{\omega^3} - i\xi' \frac{\Omega'_e{}^3}{(\omega - \vec{k}\vec{V})^3} = 1. \tag{7}$$

In order not to leave the $\Omega'_e$ out of the equation, which takes the small values, the denominators in the second and the fourth terms of equation (7) should be comparably small too. Thus the solution of equation (7) should be in the form $\omega = \vec{k}\vec{V} + \delta$ with a small $\delta$. Thus we get:

$$\frac{\Omega_e^2}{(\vec{k}\vec{V} + \delta)^2} + \frac{\Omega'_e{}^2}{\delta^2} - i\xi \frac{\Omega_e^3}{(\vec{k}\vec{V} + \delta)^3} - i\xi' \frac{\Omega'_e{}^3}{\delta^3} = 1. \tag{8}$$

Taking into account that

$$\vec{k}\vec{V} \gg \delta, \tag{9}$$

we can exclude $\delta$'s from the expression:

$$\frac{\Omega_e^2}{(\vec{k}\vec{V})^2} - i\xi \frac{\Omega_e^3}{(\vec{k}\vec{V})^3} + \frac{\Omega'^2_e}{\delta^2} - i\xi' \frac{\Omega'^3_e}{\delta^3} = 1. \quad (10)$$

Rearranging terms, we get:

$$\frac{\Omega_e^2(\vec{k}\vec{V} - i\xi\Omega_e)}{(\vec{k}\vec{V})^3} + \frac{\Omega'^2_e(\delta - i\xi'\Omega')}{\delta^3} = 1, \quad (11)$$

$$\delta^3 \Omega_e^2(\vec{k}\vec{V} - i\xi\Omega_e) + (\vec{k}\vec{V})^3 \Omega'^2_e(\delta - i\xi'\Omega'_e) =$$
$$= \delta^3(\vec{k}\vec{V})^3. \quad (12)$$

Again using the property (9), we can deduce that the term $\delta^3 \Omega_e^2(\vec{k}\vec{V} - i\xi\Omega_e)$ is much smaller than the others in the expression (12). Thus we get the algebraic equation of third degree in the canonical form:

$$\delta^3 + (-\Omega'^2_e)\delta + i\xi'\Omega'^3_e = 0. \quad (13)$$

The solution of this equation is carried out by radicals of the Cardano formula [16]:

$$\delta = \sqrt[3]{-\frac{q}{2} + \sqrt{\frac{q^2}{4} + \frac{p^3}{27}}} + \sqrt[3]{-\frac{q}{2} - \sqrt{\frac{q^2}{4} + \frac{p^3}{27}}}, \quad (14)$$

where $p = -\Omega'^2_e$ and $q = i\xi'\Omega'^3_e$. This formula can be applied if and only if the condition $\alpha\beta = -\frac{p}{3}$ is true. Indeed, taking into account that

$$\alpha = \sqrt[3]{-\frac{q}{2} + \sqrt{\frac{q^2}{4} + \frac{p^3}{27}}} =$$

$$= \sqrt[3]{-\frac{i\xi'\Omega'^3_e}{2} + \sqrt{-\frac{\xi'^2\Omega'^6_e}{4} - \frac{\Omega'^6_e}{27}}} \text{ and} \quad (15)$$

$$\beta = \sqrt[3]{-\frac{q}{2} - \sqrt{\frac{q^2}{4} + \frac{p^3}{27}}} =$$

$$= \sqrt[3]{-\frac{i\xi'\Omega'^3_e}{2} - \sqrt{-\frac{\xi'^2\Omega'^6_e}{4} - \frac{\Omega'^6_e}{27}}}, \quad (16)$$

the required criterion is met:

$$\sqrt[3]{-\frac{i\xi'\Omega'^3_e}{2} + \sqrt{-\frac{\xi'^2\Omega'^6_e}{4} - \frac{\Omega'^6_e}{27}}} \times$$

$$\times \sqrt[3]{-\frac{i\xi'\Omega'^3_e}{2} - \sqrt{-\frac{\xi'^2\Omega'^6_e}{4} - \frac{\Omega'^6_e}{27}}} = \frac{\Omega'^2_e}{3}. \quad (17)$$

And solution of the equation (13) takes the form:

$$\delta = \sqrt[3]{-\frac{i\xi'\Omega'^3_e}{2} + \sqrt{-\frac{\xi'^2\Omega'^6_e}{4} - \frac{\Omega'^6_e}{27}}} +$$
$$+ \sqrt[3]{-\frac{i\xi'\Omega'^3_e}{2} - \sqrt{-\frac{\xi'^2\Omega'^6_e}{4} - \frac{\Omega'^6_e}{27}}}. \quad (18)$$

### 3. Conclusions

The instability increment is $\gamma = -Im(\omega)$. If in a certain interval of $\vec{k}$ values, $\gamma$ is less than zero, then the perturbation increases, i.e. the environment is unstable relative to oscillations in this particular wave lengths interval. An exponential $\exp(|\gamma|t)$ increase in perturbation is meant only in the frame of the linear approximation, which is limited by the nonlinear effects in reality. In our case the expression $Im(\omega) = Im(\delta) < 0$ is true, thus the instability increment $\gamma$ is greater than zero, and the oscillations decrease with time.

Hence the system "beam – Lorentz plasma" is stable with respect to oscillations with relatively small values of $\vec{k}\vec{V}$.